\documentclass[conference]{IEEEtran}
\IEEEoverridecommandlockouts
\usepackage{cite}
\usepackage{amsmath,amssymb,amsfonts}
\usepackage{booktabs}                   
\usepackage{multicol}                   
\usepackage{multirow}                   
\usepackage{makecell}
\usepackage{algorithmic}
\usepackage{graphicx}
\usepackage{textcomp}
\usepackage{xcolor}
\def\BibTeX{{\rm B\kern-.05em{\sc i\kern-.025em b}\kern-.08em
    T\kern-.1667em\lower.7ex\hbox{E}\kern-.125emX}}
        
\usepackage{fancyhdr}
\begin{document}

\title{The OMG-Empathy Dataset: Evaluating the Impact of Affective Behavior in Storytelling\\}

\author{\parbox{16cm}{\centering
{{\large Pablo~Barros$^1$, Nikhil~Churamani$^{3}$, Angelica~Lim$^2$ and Stefan~Wermter$^1$}\\}
    {\normalsize
    \vspace{1mm}
    {$^1$ Knowledge Technology, Department of Informatics, University of Hamburg, Hamburg, Germany\\ E-mail:~\{barros, wermter\}@informatik.uni-hamburg.de\\}
    {$^2$ School of Computing Science, Simon Fraser University, Burnaby, BC, Canada\\ E-mail:~angelica@sfu.ca\\}
    {$^3$ Department of Computer Science and Technology, University of Cambridge, Cambridge, UK\\E-mail:~Nikhil.Churamani@cl.cam.ac.uk\\}
    }}
}

\maketitle
\thispagestyle{fancy}

\begin{abstract}

Processing human affective behavior is important for developing intelligent agents that interact with humans in complex interaction scenarios. A large number of current approaches that address this problem focus on classifying emotion expressions by grouping them into known categories. Such strategies neglect, among other aspects, the impact of the affective responses from an individual on their interaction partner thus ignoring how people empathize towards each other. This is also reflected in the datasets used to train models for affective processing tasks. Most of the recent datasets, in particular, the ones which capture natural interactions (``in-the-wild'' datasets), are designed, collected, and annotated based on the recognition of displayed affective reactions, ignoring how these displayed or expressed emotions are perceived. In this paper, we propose a novel dataset composed of dyadic interactions designed, collected and annotated with a focus on measuring the affective impact that  eight different stories have on the \textit{listener}. Each video of the dataset contains around 5 minutes of interaction where a speaker tells a story to a \textit{listener}. After each interaction, the \textit{listener} annotated, using a valence scale, how the story impacted their affective state, reflecting how they empathized with the speaker as well as the story. We also propose different evaluation protocols and a baseline that encourages participation in the advancement of the field of artificial empathy and emotion contagion.
\end{abstract}

\begin{IEEEkeywords}
Empathy, Dyadic Interactions, Affective Behaviour
\end{IEEEkeywords}

\section{Introduction}

The recent increased interest in Affective Computing~\cite{Picard1997AC} has resulted in effective emotion expression recognition solutions~\cite{Soleymani2017}. However, the inclusion of affective understanding in the decision-making process of an agent goes beyond expression perception. Treating the perception of affect from expressions as the goal minimizes the contribution of emotions in complex interaction scenarios~\cite{Heberlein2009}. To create a general model for affect that can be used as a modulator for learning different cognitive tasks, such as modeling intrinsic motivation, creativity, dialog processing, grounded learning, and human-like communication, only affective perception cannot be the pivotal focus. The integration of emotion perception with intrinsic concepts of affect understanding, such as empathy, is required to model the necessary complexity of interaction and realize adaptability in an agent's social behavior~\cite{Venkatesh2000}.

One of the most important aspects of affective understanding in humans is the ability to understand and develop empathetic behavior. Empathy is usually associated with cognitive behavior, which has its roots in the developmental aspects of human communication~\cite{Decety2009}. It helps us to promote natural communication by transferring affective behavior from others to our intrinsic affective state and can be understood as the impact that an emotional situation has on a person's affective state. In this sense, the contextual situation of interaction is one of the most important factors in developing empathy~\cite{Melloni2014}. Understanding why, and how, the other person demonstrates an affective behavior helps us to develop an embodied interaction with them. 

Embedding empathetic understanding in an artificial agent gives it the ability to use the contextual perception of an interaction to modulate its intrinsic affective state~\cite{Asada2015}. Recent approaches propose to embed artificial empathy in robots and allow them to be used in close-to-real-world scenarios~\cite{Leite2013, Giannopulu2018}. Such models, although based on different psychological aspects of empathy, make use of very controlled environments, and thus, are not suited for unconstrained and real-world scenarios. Different from affective recognition problems, empathy relies mostly on contextual, personalized and continuous interactions. By bonding with one specific person over time, we can develop a specific empathetic response towards that person~\cite{Rossi2017}. 

To encourage the development of artificial empathy models which are suitable to be used in real-world scenarios, we propose the OMG-Empathy Dataset, along with two different evaluation protocols. The dataset presents a realistic approach for training and evaluating artificial empathy systems for real-world applications, focusing on the impact that an affective interaction has on a \textit{listener}. It is composed of $7$ hours of audio-visual recordings of human-human interactions, collected with $10$ different participants interacting with $4$ different \textit{speakers}. Each participant held $2$ dialogues with each \textit{speaker}, each of them based on a different storyline. Each story detailed a specific fictional situation and it demanded gradual changes in affective behavior from the \textit{speaker}. With $8$ different stories per participant, a total of $80$ different interaction videos were recorded with each video spanning on average $5$ minutes and $12$ seconds, providing us with $415$ minutes (around $7$ hours) of recordings. 

Immediately after each session, the participants were asked to watch the interaction again to recall and annotate their intrinsic affective state during the interaction using a valence scale ranging from negative to positive values. The annotation was recorded continuously with the help of a joystick to assure that an accurate and fluid account of the listener's emotional state, while the \textit{speaker} tells the story, can be recorded.  

The annotation strategy forms the basis of the proposed evaluation protocols and supports research in artificial empathy at two levels, namely, \textit{personalized} empathy and \textit{generalized} empathy. In the \textit{Personalized} empathy protocol, we want to evaluate how different models can learn the emotional impact that the stories have on a person-specific scenario. The \textit{Generalized} empathy protocol, on the other hand, evaluates the ability of the proposed solution to model the aggregated emotional impact within all \textit{speaker} corresponding to one specific story.


Besides the design and collection strategies, we also provide a broader analysis of the annotation distribution. The analysis illustrates how our self-assessment annotations are distributed over all the stories and individual persons. Finally, we propose a baseline for both protocols based on a deep neural network which processes the audio and visual stimuli from both \textit{listener} and \textit{speakers}. We hope that our dataset, the analysis, and the proposed protocols boost the development of real-world applications for artificial agents dealing with empathy.

\section{Beyond Affect Recognition}


Empathy plays an important role in the development, perception, and understanding of social interactions. It is explained as the basis of how we understand each other~\cite{Smith2017}. Empathetic perception and behavior enable humans to form stronger social bonds and improve collaboration in different tasks~\cite{Howe2017}. Adapting such empathetic mechanisms in robots and similar interactive agents allows them to be more than just emotional expression recognition machines~\cite{Cramer2010}. Although highly desired, research towards realizing this adaptation in artificial agents is still far behind when compared to emotion perception research. 

One of the main problems faced while designing empathetic agents is to model the subjective complexities of empathy into computational models. Most of the recent applications of empathy in robots, for instance, are hardly distinguishable from simple imitation mechanisms~\cite{Paiva2017}. Such models are related to the concept of emotional contagion~\cite{Hatfield1993} which explains how humans share their emotional states with others by imitating their emotional state. Although emotional contagion is an important mechanism to strengthen social connections within a certain contextual event~\cite{Barsede2002}, it is still not enough for modeling empathy~\cite{DeWaal2008} and, specifically, the impact of an affective interaction. For example, while telling a happy story, a storyteller would be much more engaged with the listener when both share the same affective states through the story. However, to share the same affective state of the storyteller, the listener has to be impacted by what was said, and by how the storyteller told the story. A computational model for emotional contagion, mostly based on recognizing and imitating the storyteller emotion expressions, would be sufficient to emulate the behavior of the storyteller. To model the listener behavior, however, focusing only on the affective behavior of the storyteller would not be enough. Such a model would need to process the perceived affect, the contextual information of the story, and how this, in particular, affects that particular listener.


One of the bottlenecks of modeling the impact of such scenarios on the listener is how to train and evaluate such computational models. With the recent interest in emotion perception, several different datasets have been published in recent years~\cite{lucey2010extended, zadeh2016mosi, barros2018omg, Dhall2018}. These datasets focus on different perspectives of emotion perception and include a wide range of characteristics from \textit{in-the-wild} multimodal data, to \textit{controlled} and \textit{induced} emotional reactions. The most recent solutions for emotion recognition make use of such different conditions to achieve impressive performance on instantaneous emotion recognition. However, most of these data-driven solutions are focused on recognizing or describing the emotional state of a single person over a single instance of emotional display. They are also annotated by external evaluators which focus on how the persons are expressing emotions. They are suitable for empathetic models based on emotional contagion but fail to provide a standardized platform for training and evaluating empathetic behavior based on the impact of a perceived affect. 

There also exist several corpora which focus on continuous dialogues, mostly dyadic interactions~\cite{busso2017msp,McKeown2012,Avec2012}. The possibility of extracting long-term contextual information makes these datasets suitable for long-term emotion recognition. In recent years, different computational solutions for emotion recognition on dialogues based on hand-crafted feature extractions~\cite{Jin2015} and using deep neural models~\cite{Sarma2018} are trained and evaluated in such datasets. Some of these use contextual information to provide a general emotional description of the scenes~\cite{Majumder2018, Hazarika2018}. Although they provide contextual information, such datasets still are mainly used for training models for the recognition of affective display. When using these datasets, it is not possible to model, and subsequently evaluate, the impact that the conversation had on the affective state of the participants. 

Most of the deep learning solutions which are trained on such datasets would fall short in modeling the affective impact of the interaction within the subjects. The recent models which attempt to do so only partially benefit from such datasets. Such models make use of emotion recognition to provide imitation-based reactions~\cite{Boucenna2014}, simple threshold-based decision-making scenarios~\cite{Ranieri2016} or even affective memory development~\cite{Barros2017}. Having a complementary dataset to learn the empathetic behavior of the subjects would benefit such models greatly.

\begin{table*}[t]
\centering
\caption{Topics for the eight stories told by the speakers and the encoded emotional state in the stories.}
\setlength\tabcolsep{3.0pt}

\small{
\begin{tabular}{ c| l | l | c | l | l}
    \toprule
    \textbf{Story}             & \makecell[c]{\textbf{Topic}}         &\makecell[c]{\textbf{Emotional State}}  & \textbf{Story}             & \makecell[c]{\textbf{Topic}}         &\makecell[c]{\textbf{Emotional State}}         \\ \midrule
    $1$ & \makecell[l]{I miss my childhood friend.}             & Sadness, Nostalgia      & $5$ & \makecell[l]{I had an adventurous travelling experience.}   & Surprise, Excitement    \\ \midrule
    $2$ & How I started a band!                         & Happiness, Excitement   & $6$ & I cheated on an exam when I was younger.      & Sadness, Shame          \\ \midrule
    $3$ & My relation with my old dog.                      & Sadness, Grief          & $7$ & I won a martial arts challenge.               & Happiness, Pride        \\ \midrule
    $4$ & I had a bad flight experience.                & Fear, Panic             & $8$ & I ate a very bad food item.                   & Disgust, Shame          \\ \bottomrule
    
  \end{tabular} 
}
\label{tab:stories}
\end{table*}

\section{The OMG-Empathy Prediction Dataset}

The OMG-Empathy dataset\footnote{https://bit.ly/2SL4mLC} is designed to provide a basis for models that aim to predict how affective interactions impact different individuals. The dataset consists of recordings of semi-scripted dyadic interactions between a \textit{speaker} and a \textit{listener} discussing a specific topic where the \textit{speaker} leads the conversation. The speaker tells a fictional story about their recent encounters while the listener reacts to their story, empathizing with them. Eight topics (see \textit{Table~\ref{tab:stories}}) were created for the \textit{speaker} to talk about, each of them corresponding to one or more emotional state. The \textit{listeners} were not informed that the topics were fictional.

\begin{table}[t]
\centering
\caption{Information about the different speakers telling stoties to the participants.}
\small{
\begin{tabular}{ c| c | c | c }
    \toprule
    \textbf{Speaker}    & \textbf{Stories}  &\textbf{ Mother Tongue}    &  \textbf{Style}   \\ \midrule
    Speaker $1$         & $1,2$             & English                   & Introverted       \\ \midrule
    Speaker $2$         & $3,4$             & Hindi                     & Calm              \\ \midrule
    Speaker $3$         & $5,6$             & Portuguese                & Extroverted       \\ \midrule
    Speaker $4$         & $7,8$             & Italian                   & Excited           \\ \bottomrule
  \end{tabular} 
}
\label{tab:omgActor}
\end{table}

The \textit{speakers} were free and encouraged to improvise on each of these topics so that we recorded a natural conversation scenario but were instructed to maintain control over the conversation. This way we guaranteed that the recorded interactions were not completely one-sided but at the same time that the \textit{listener} did not take over the direction of the conversation. A total of $4$ different \textit{speakers} interacted with all the participants, each of them telling $2$ different stories to each participant. Each \textit{speaker} was recruited from the departmental staff and presented a different style of storytelling. The styles were pre-defined and followed $4$ different personality traits, namely, \textit{introverted}, \textit{calm}, \textit{extroverted} and \textit{excited}. The \textit{speaker} responsible for the introverted style presented the stories in a very monotonic manner, avoiding much eye contact with the \textit{listener}. The \textit{speaker} with the calm style told the stories in a normal voice tone, maintaining minimum interaction, while the \textit{speaker} telling the stories in an extroverted manner made more interactions with the participants, as well as presented the stories using a higher activation on its emotion expressions. Finally, the excited \textit{speaker} presented the stories in an over-reactive way, making use of a lot of gesticulations and different facial expressions. Each speaker comes from a different country, but all were able to speak English fluently. \textit{Table~\ref{tab:omgActor}} records details about each speaker, the stories narrated by them and their interaction style.

The \textit{speakers} were asked to narrate the story following a pre-defined set of key events but were free to improvise and to tell the stories in their own way. This resulted in the same speaker telling the story slightly differently for each participant but maintaining the sequence of key events in each story in a similar storytelling style.
 
\begin{figure}
\centering
\includegraphics[width=0.48\textwidth]{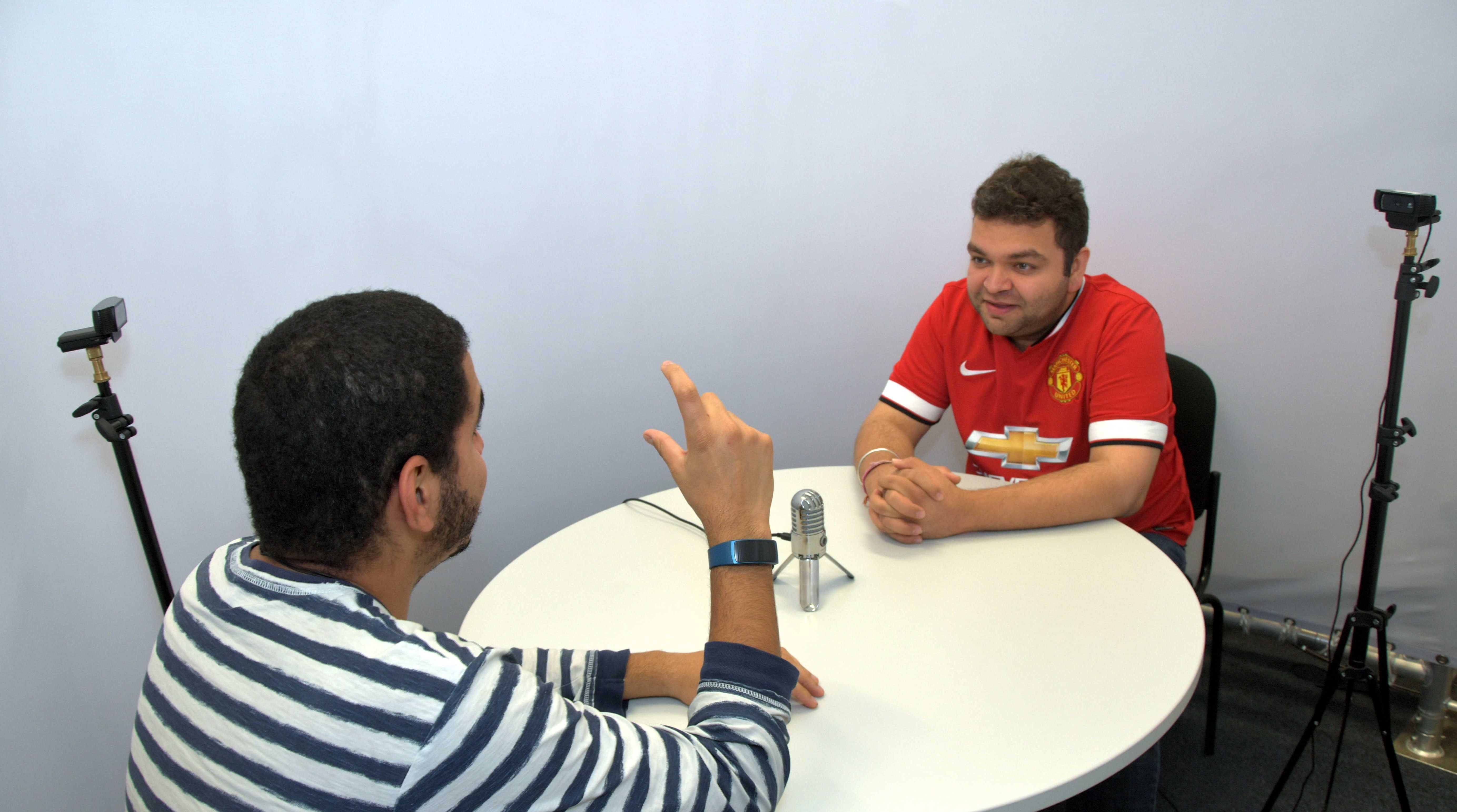}
\caption{\textit{Speaker} (\textit{left}) interacting with the \textit{listener} (\textit{right}).}
\label{fig:scenes}
\end{figure}
\subsection{Data Collection}

We recorded the audio and visual data from both the \textit{speakers} and \textit{listener} for each interaction. The \textit{speaker} and the \textit{listener} were seated in front of each other. Two cameras recorded the upper-body for each of them and a microphone placed in the center of the table recorded the whole conversation. Figure~\ref{fig:scenes} illustrates the recording scenario.

Each \textit{listener} had two sessions of recordings, on separate days. Each session lasted $45$ minutes. In the first session, the \textit{listener} interacted with $2$ \textit{speakers}, each of them telling $2$ stories. The order of stories and \textit{speakers} were the same for all the \textit{listeners}, so if any bias was introduced its impact was contained.


We had a total of $10$ \textit{listeners}, each one taking part in all the $8$ interactions (stories), as detailed in Table~\ref{tab:omgSubject}. Each \textit{listener} came from a different nationality with Germany being the only country which is repeated. The dataset is also gender-balanced, having $5$ female and $5$ male \textit{listeners}. The variation in the cultural background of the \textit{listeners} in the dataset imposes a challenge on predicting the impact of the affective behavior of the \textit{speaker}, but also an opportunity for modeling different aspects of how the dialogues impact different persons.

\begin{figure}
\centering
\includegraphics[width=0.48\textwidth]{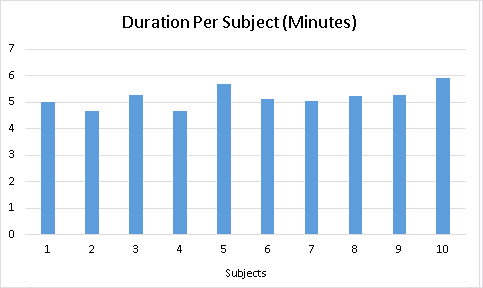}
\caption{Average duration, in minutes, of the videos for each \textit{listener}.}
\label{fig:durationSubject}
\end{figure}

 Each of the $80$ recorded videos spanned for an average of $6$ minutes and $12$ seconds, providing us with $480$ minutes (around $8$ hours) of recordings. While interacting with different \textit{listeners}, the \textit{speakers} extended or reduced the dialogue duration spontaneously. Figure~\ref{fig:durationSubject} illustrates the average duration of the interactions per \textit{listener}. 

\begin{table}
\caption{Summary information about the listeners of our experiments.}
    \center 
    \small{
    \begin{tabular}{c| c | c | c} \toprule
    \textbf{Listeners} & \textbf{Mother Tongue} &  \textbf{Gender} &\textbf{ Age Group} \\ \midrule
    Listener $1$ & Chinese       &   Female  & $19-30$ \\ \hline
    Listener $2$ & German        &   Male    & $19-30$ \\ \hline
    Listener $3$ & Farsi         &   Male    & $19-30$ \\ \hline
    Listener $4$ & Arabic        &   Male    & $19-30$ \\ \hline
    Listener $5$ & Chinese       &   Female  & $19-30$ \\ \hline
    Listener $6$ & Albanian      &   Female  & $19-30$ \\ \hline
    Listener $7$ & Greek         &   Female  & $19-30$ \\ \hline
    Listener $8$ & German        &   Male    & $19-30$ \\ \hline
    Listener $9$ & Portuguese    &   Male    & $19-30$ \\ \hline
    Listener $10$ & Urdu         &   Female  & $19-30$ \\ \bottomrule
  \end{tabular} 
  }
\label{tab:omgSubject}
\end{table}

\begin{figure}
\centering
\includegraphics[width=0.48\textwidth]{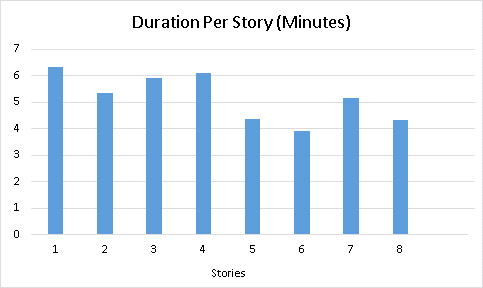}
\caption{Average duration, in minutes, of the videos for each story.}
\label{fig:durationStory}
\end{figure}

The variation in the interaction duration can also be seen in the average duration per story, as illustrated in Figure~\ref{fig:durationStory}. While story $1$ lasted, on average, for more than $6$ minutes, story $6$ only lasted on average for about $4$ minutes. 

The average duration per video gives us an insight on how specific \textit{listeners} behave and also on how different stories impact each \textit{listener}. While \textit{listener} $4$ seems to prefer short interactions, \textit{listener} $10$ had more engagement during the dialogues. 



The participants were recruited from the institution using mailing lists and on-campus recruitment. Each \textit{listener} was fully informed about the goal of the data collection and provided informed consent, giving permission to have their data publicly available. The consent form and the experimental protocol was approved by the ethics committee of the University of Hamburg.

\subsection {Self-assessment Annotation}

Immediately after each recording session, we asked the \textit{listeners} to watch the interactions again on a computer screen and use a joystick to annotate how the interaction impacted his affective state in terms of valence using a continuous scale ranging from positive $(1)$ to negative $(-1)$ values. The use of the joystick allowed for continuous and gradual tracking of annotations which are temporally related to the interaction scenario.

\begin{figure}
\centering
\includegraphics[width=0.4\textwidth]{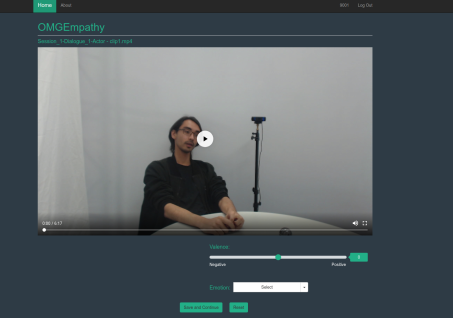}
\caption{The User Interface of the tool used for the self-assessment annotations.}
\label{fig:ktAnnotationTool}
\end{figure}

To annotate the videos, the \textit{listeners} used a modified version of the KT-Annotation Tool~\cite{barros2018omg} which was designed as a dynamic tool for collecting dataset annotations. The tool provides annotators with a web-based system that can be adjusted for different annotation scenarios. It was developed using the Django\footnote{https://www.djangoproject.com [Accessed 28.03.2019]} framework with a secure back-end built using SQLite\footnote{https://sqlite.org/ [Accessed 28.09.2018]}. We modified the tool adding joystick (Gamepad\footnote{https://github.com/neogeek/gamepad.js [Accessed 28.03.2019]}) support to make use of an analog joystick which was used by the \textit{listeners} to annotate their self-assessment feeling. Figure~\ref{fig:ktAnnotationTool} illustrates the tool interface that was developed for this project. 

\section{Data Post-processing}

Once all the videos were recorded, they had to be synchronized, cleaned and matched with the annotations. We synced the videos based on the audio information captured from each camera. It is important that the \textit{listener} and the \textit{speaker} videos are frame-by-frame precisely synchronized, so they correlate to the annotations.

\begin{figure}
\centering
\includegraphics[width=0.48\textwidth]{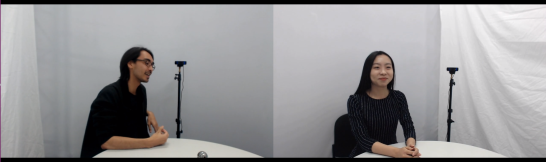}
\caption{Example of one synced video with the \textit{speaker} (on the left) and the \textit{listener} (on the right).}
\label{fig:exampleVideo}
\end{figure}

We stitched each \textit{speaker} and \textit{listener} video pair into one single video, as illustrated in Figure~\ref{fig:exampleVideo}. This facilitates the distribution of the data and guarantees that the videos are synced frame-wise. Each video has a resolution of $2560\times720$, with a frame-rate of $25$ frames-per-second and an audio sample rate of $44100$ Hz. We standardize all the videos to make sure that the \textit{speaker} is always on the left, and the \textit{listener} is always on the right.

The annotations were collected continuously over the duration of the video. After the annotations were collected, we re-sample them using a windowed-averaging approach resulting in one annotation per video frame. The annotations are provided as \textit{.csv} files, one file per video. Each row of the file corresponds to one annotation corresponding to one frame of the video.

\section{Experimental Protocol}

To provide a standardized method to evaluate our dataset, we propose here two different protocols: a personalized and a generalized one. 

For both protocols, the dataset has pre-defined separation sets: training, validation, and testing. We separate our samples based on balancing the training and testing sets based on the self-assessment annotations. From all the stories, we set $4$ of them for training, $1$ for validation and $3$ for testing. Each story has $10$ videos associated with it, one for each \textit{listener}. 

We selected stories $2$, $4$, $5$ and $8$ for the training subset, with a total of $211$ minutes of video data. Stories $3$, $6$ and $7$ were selected for the testing subset, totaling $170$ minutes of video data. And finally, story $1$ was selected for the validation subset, totaling $73$ minutes of video information.

\subsection{Personalized Empathy}

\begin{figure*}
\centering
 {\includegraphics[width=0.35\textwidth]{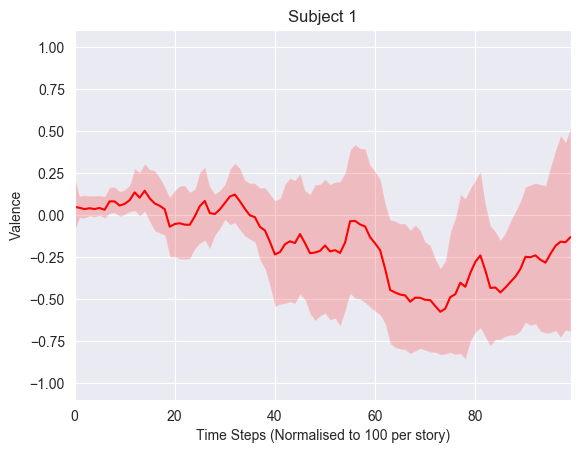}}
 {\includegraphics[width=0.35\textwidth]{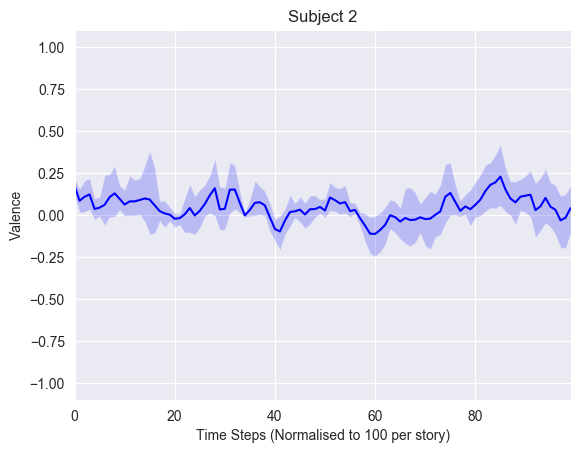}}

   \caption{Self-annotation for \textit{listeners} 1 and 2 across all the $8$ stories, re-sampled to a $100\%$ scale representing the total video duration.}
 \label{fig:subject}
\end{figure*}

The Personalized Empathy protocol focuses on modeling the impact that all the stories would have on the affective state of a specific person. It evaluates the ability of proposed models to learn the empathetic impact on each of the \textit{listeners} over a newly perceived story. Each person is impacted differently by the specific stories, and the proposed models must consider this. Figure~\ref{fig:subject} illustrates an example of this behavior by plotting the self-assessments of \textit{listeners} 1 and 2 re-sampled to a $100\%$ scale representing the total video duration. While \textit{Listener} $2$ demonstrates a very steady behavior over all the stories, \textit{Listener} $1$ presents a wider range of valence variation. 

\subsection{Generalized Empathy}

\begin{figure*}
\centering
 {\includegraphics[width=0.35\textwidth]{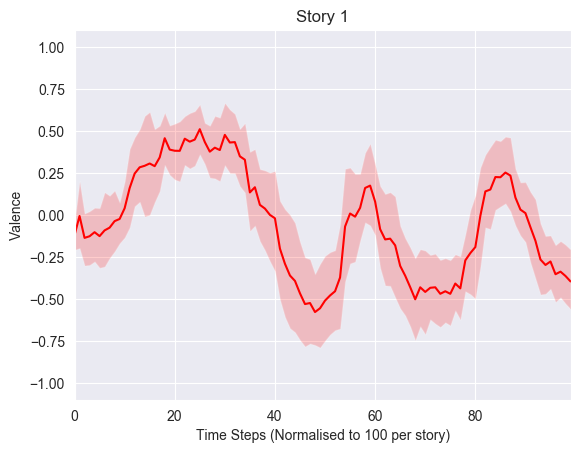}}
 {\includegraphics[width=0.35\textwidth]{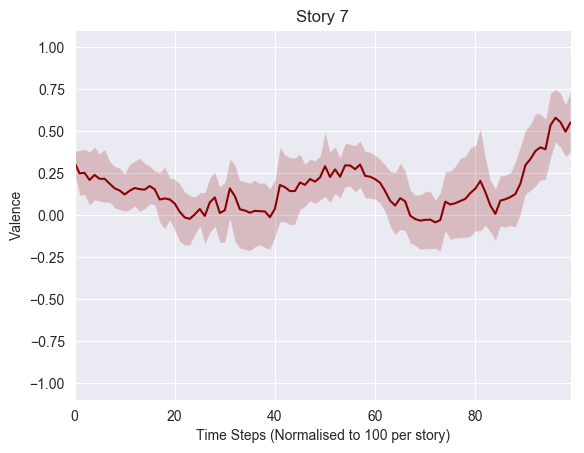}}

   \caption{Self-annotation for the average of all \textit{listeners} for stories 1 and 7 re-sampled to a $100\%$ scale representing the total video duration.}
 \label{fig:story}
\end{figure*}

The Generalized Empathy protocol focuses on the prediction of the general impact each of the stories had over of all the \textit{listeners}. This is obtained by averaging over all the valences of the \textit{listeners} over one specific story. We illustrate the difference between the stories by showing in Figure~\ref{fig:story} the re-sampled self-annotations, on a $100\%$ scale representing the video duration, for Story $1$ and Story $7$. While Story $1$ presents a wider fluctuation on the general valence, caused by the nostalgic content of the story, Story $7$ has a more stable annotation as it relates to a happier and more exciting story.

For this protocol, we encourage the development of models to take into consideration the aggregated behavior of all the participants for each story and to generalize this behavior in a newly perceived story. 

\subsection{Evaluation Metrics}

To have an adequate and reproducible measure for each of the protocols we use the Concordance Correlation Coefficient (CCC)~\cite{Lawrence1989} as an objective evaluation metric. It measures the similarity between the predictions of a model and the \textit{listener`s} own assessment. The CCC can be computed as:

\begin{equation}
CCC = \frac{2 \rho \sigma_x \sigma_y}{\sigma_{x}^2 + \sigma_{y}^2 + (\mu_x - \mu_y)^2}
\label{eq:ccc}
\end{equation}
where $\rho$ is the Pearson's Correlation Coefficient between model predictions labels and the annotations, $\mu_x$ and $\mu_y$ denote the mean for model predictions and the annotations and $\sigma_{x}^2$ and $\sigma_{y}^2$ are the corresponding variances.

For the Personalized Empathy protocol, the CCC is calculated between the output of a certain computational model and each of the \textit{listeners's} own assessment for each of the stories. Each \textit{listener} will have one CCC measure averaged over all the stories.

The Generalized Empathy track evaluates the CCC between the output of a certain computational model and the self-assessment of each \textit{listener} over all the stories. Each story will have one CCC measure, calculated as the average over all the listeners.

\section{Baseline and Results}

As a baseline for both protocols, we decided to adapt a deep neural network for representing the multimodal stimuli from both \textit{listener} and \textit{speaker}. To provide a competitive baseline, we decided to adapt the winner model of the recent OMG-Emotion Recognition challenge~\cite{Zheng2018}. This model proposed a multi-channel convolution neural network for multimodal emotion recognition based on a temporal attention layer to provide the recognition of expressions over time. 

The baseline model is composed of two individual convolution channels, one for extracting features from faces and one to extract the features from speech signals. The face expression channel is based on the VGG16~\cite{Simonyan2014} architecture and is connected to a Long-Short Memory (LSTM) layer with $256$ hidden units to extract spatial-temporal features from a sequence of frames. In our baseline, we extract the faces from each frame using the Dlib~\cite{King2009} framework. The auditory channel is created based on the same topology as the SoundNet~\cite{Soundnet2016}, which uses $1D$ convolutions to extract information from raw audio waves. These two channels are trained individually, and after training their output are concatenated and used to train a Support Vector Machine (SVM).

To explore all the perception aspects that the dataset provides, we train and evaluate the baseline model based on the stimuli coming only from the \textit{speaker}, only the listener, and a late-fusion concatenation of both, by using the extracted features of both to train the SVM. We then calculate the CCC between the perception models and the self-assessment annotations using both proposed evaluation protocols. The results can be found in Table~\ref{tab:baselineResult}.

\begin{table}
\caption{Resulting CCC between the self-assessment and the output of a perception model on describing the \textit{speaker}, the \textit{listener} and a combination of both.}
    \center 
    \small{
    \begin{tabular}{c | c| c} \toprule
    \textbf{Observation} & \textbf{Personalized Score} & \textbf{Generalized Score} \\ \midrule
    \textit{Speaker}       & $0.11$    & $0.13$    \\ \hline
    \textit{Listener}    & $0.19$    & $0.23$    \\\hline
    Both        & $0.17$    & $0.19$    \\\bottomrule
  \end{tabular}
  }
\label{tab:baselineResult}
\end{table}

Although the model presents state-of-the-art performance for emotion recognition tasks, it's performance on recognizing the impact of the stories on the \textit{listener's} affective state is poor. This is in agreement with the hypothesis that for processing empathy in such a real-world scenario, more complex models are necessary~\cite{lim2015recipe}. Solutions which take into consideration contextual processing, and most importantly, which can generalize the individual impact assessment of each \textit{listener} towards the stories are expected to provide better performance.

\section{Discussions and Conclusions}

This dataset presents a novel mechanism for training and evaluating computational models to predict the impact that affective interactions have on different \textit{listeners}. It contributes to the artificial empathy community by introducing two standard evaluation protocols for assessing the emotional impact of the stories with very objective measures.

The experimental design and data collection are performed to provide variability on the impact of the stories to the \textit{listeners}. Yet, we keep a controllable and reproducible environment within the stories so the \textit{listeners'} self-assessments have a meaningful representation. By analyzing the annotations, we can validate that each story has a different impact on the  \textit{listeners'} overall affective state. At the same time all the \textit{listeners} reported a similar impact behavior over similar stories.

Of course, keeping such controllable scenario comes with costs: there might be a disassociation between the \textit{speakers'} pre-defined stories and the way they would express real stories which we did not take into consideration. We also did not investigate the relationship between the \textit{listeners'} assessment and the real cause of the annotated impact. We provided, however, the first steps towards these discussions.

Our baseline experiments demonstrate that, although automatic emotion expression recognition has achieved impressive levels of performance in recent years, it is still not performant enough for empathetic modeling. We expect that models which take into consideration the contextual information of the videos would outperform the proposed baseline. Also, models which are able to learn personalized representations of intrinsic emotions would have an improved performance on this dataset.

\section*{Acknowledgment}

The authors gratefully acknowledge partial support from the German Research Foundation DFG under project CML (TRR 169). This work was completed when Nikhil Churamani was with Knowledge Technology, University of Hamburg.

\bibliographystyle{IEEEtran}
\bibliography{bib}

\begin{thebibliography}{10}
\providecommand{\url}[1]{#1}
\csname url@samestyle\endcsname
\providecommand{\newblock}{\relax}
\providecommand{\bibinfo}[2]{#2}
\providecommand{\BIBentrySTDinterwordspacing}{\spaceskip=0pt\relax}
\providecommand{\BIBentryALTinterwordstretchfactor}{4}
\providecommand{\BIBentryALTinterwordspacing}{\spaceskip=\fontdimen2\font plus
\BIBentryALTinterwordstretchfactor\fontdimen3\font minus
  \fontdimen4\font\relax}
\providecommand{\BIBforeignlanguage}[2]{{%
\expandafter\ifx\csname l@#1\endcsname\relax
\typeout{** WARNING: IEEEtran.bst: No hyphenation pattern has been}%
\typeout{** loaded for the language `#1'. Using the pattern for}%
\typeout{** the default language instead.}%
\else
\language=\csname l@#1\endcsname
\fi
#2}}
\providecommand{\BIBdecl}{\relax}
\BIBdecl

\bibitem{Picard1997AC}
R.~W. Picard, \emph{Affective Computing}.\hskip 1em plus 0.5em minus
  0.4em\relax Cambridge, MA, USA: MIT Press, 1997.

\bibitem{Soleymani2017}
M.~Soleymani, D.~Garcia, B.~Jou, B.~Schuller, S.-F. Chang, and M.~Pantic, ``A
  survey of multimodal sentiment analysis,'' \emph{Image and Vision Computing},
  vol.~65, pp. 3--14, 2017.

\bibitem{Heberlein2009}
A.~S. Heberlein and A.~P. Atkinson, ``Neuroscientific evidence for simulation
  and shared substrates in emotion recognition: beyond faces,'' \emph{Emotion
  Review}, vol.~1, no.~2, pp. 162--177, 2009.

\bibitem{Venkatesh2000}
V.~Venkatesh, ``Determinants of perceived ease of use: Integrating control,
  intrinsic motivation, and emotion into the technology acceptance model,''
  \emph{Information systems research}, vol.~11, no.~4, pp. 342--365, 2000.

\bibitem{Decety2009}
J.~E. Decety and W.~E. Ickes, \emph{The social neuroscience of empathy.}\hskip
  1em plus 0.5em minus 0.4em\relax MIT Press, 2009.

\bibitem{Melloni2014}
M.~Melloni, V.~Lopez, and A.~Ibanez, ``Empathy and contextual social
  cognition,'' \emph{Cognitive, Affective, \& Behavioral Neuroscience},
  vol.~14, no.~1, pp. 407--425, 2014.

\bibitem{Asada2015}
M.~Asada, ``Towards artificial empathy,'' \emph{International Journal of Social
  Robotics}, vol.~7, no.~1, pp. 19--33, 2015.

\bibitem{Leite2013}
I.~Leite, A.~Pereira, S.~Mascarenhas, C.~Martinho, R.~Prada, and A.~Paiva,
  ``The influence of empathy in human--robot relations,'' \emph{International
  journal of human-computer studies}, vol.~71, no.~3, pp. 250--260, 2013.

\bibitem{Giannopulu2018}
I.~Giannopulu, K.~Terada, and T.~Watanabe, ``Emotional empathy as a mechanism
  of synchronisation in child-robot interaction,'' \emph{Frontiers in
  Psychology}, vol.~9, p. 1852, 2018.

\bibitem{Rossi2017}
S.~Rossi, F.~Ferland, and A.~Tapus, ``User profiling and behavioral adaptation
  for hri: A survey,'' \emph{Pattern Recognition Letters}, vol.~99, pp. 3--12,
  2017.

\bibitem{Smith2017}
J.~Smith, ``What is empathy for?'' \emph{Synthese}, vol. 194, no.~3, pp.
  709--722, 2017.

\bibitem{Howe2017}
D.~Howe \emph{et~al.}, ``Empathy, social intelligence and relationship-based
  social work,'' \emph{Zeszyty Pracy Socjalnej}, vol. 2017, no. numer 1, pp.
  1--12, 2017.

\bibitem{Cramer2010}
H.~Cramer, J.~Goddijn, B.~Wielinga, and V.~Evers, ``Effects of (in) accurate
  empathy and situational valence on attitudes towards robots,'' in
  \emph{Human-Robot Interaction (HRI), 2010 5th ACM/IEEE International
  Conference on}.\hskip 1em plus 0.5em minus 0.4em\relax IEEE, 2010, pp.
  141--142.

\bibitem{Paiva2017}
A.~Paiva, I.~Leite, H.~Boukricha, and I.~Wachsmuth, ``Empathy in virtual agents
  and robots: a survey,'' \emph{ACM Transactions on Interactive Intelligent
  Systems (TiiS)}, vol.~7, no.~3, p.~11, 2017.

\bibitem{Hatfield1993}
E.~Hatfield, J.~T. Cacioppo, and R.~L. Rapson, ``Emotional contagion,''
  \emph{Current directions in psychological science}, vol.~2, no.~3, pp.
  96--100, 1993.

\bibitem{Barsede2002}
S.~G. Barsade, ``The ripple effect: Emotional contagion and its influence on
  group behavior,'' \emph{Administrative Science Quarterly}, vol.~47, no.~4,
  pp. 644--675, 2002.

\bibitem{DeWaal2008}
F.~B. De~Waal, ``Putting the altruism back into altruism: the evolution of
  empathy,'' \emph{Annu. Rev. Psychol.}, vol.~59, pp. 279--300, 2008.

\bibitem{lucey2010extended}
P.~Lucey, J.~F. Cohn, T.~Kanade, J.~Saragih, Z.~Ambadar, and I.~Matthews, ``The
  extended cohn-kanade dataset (ck+): A complete dataset for action unit and
  emotion-specified expression,'' in \emph{Computer Vision and Pattern
  Recognition Workshops (CVPRW), 2010 IEEE Computer Society Conference
  on}.\hskip 1em plus 0.5em minus 0.4em\relax IEEE, 2010, pp. 94--101.

\bibitem{zadeh2016mosi}
A.~Zadeh, R.~Zellers, E.~Pincus, and L.-P. Morency, ``Mosi: multimodal corpus
  of sentiment intensity and subjectivity analysis in online opinion videos,''
  \emph{arXiv preprint arXiv:1606.06259}, 2016.

\bibitem{barros2018omg}
P.~Barros, N.~Churamani, E.~Lakomkin, H.~Sequeira, A.~Sutherland, and
  S.~Wermter, ``{The OMG-Emotion Behavior Dataset},'' in \emph{Proceedings of
  the International Joint Conference on Neural Networks (IJCNN)}.\hskip 1em
  plus 0.5em minus 0.4em\relax IEEE, 2018, pp. 1408--1414.

\bibitem{Dhall2018}
A.~Dhall, A.~Kaur, R.~Goecke, and T.~Gedeon, ``Emotiw 2018: Audio-video,
  student engagement and group-level affect prediction,'' in \emph{Proceedings
  of the 2018 on International Conference on Multimodal Interaction}, ser. ICMI
  '18.\hskip 1em plus 0.5em minus 0.4em\relax New York, NY, USA: ACM, 2018, pp.
  653--656.

\bibitem{busso2017msp}
C.~Busso, S.~Parthasarathy, A.~Burmania, M.~AbdelWahab, N.~Sadoughi, and E.~M.
  Provost, ``Msp-improv: An acted corpus of dyadic interactions to study
  emotion perception,'' \emph{IEEE Transactions on Affective Computing}, no.~1,
  pp. 67--80, 2017.

\bibitem{McKeown2012}
G.~McKeown, M.~Valstar, R.~Cowie, M.~Pantic, and M.~Schroder, ``The semaine
  database: Annotated multimodal records of emotionally colored conversations
  between a person and a limited agent,'' \emph{IEEE Transactions on Affective
  Computing}, vol.~3, no.~1, pp. 5--17, 2012.

\bibitem{Avec2012}
B.~Schuller, M.~Valstar, R.~Cowie, and M.~Pantic, ``Avec 2012: the continuous
  audio/visual emotion challenge-an introduction,'' in \emph{Proceedings of the
  14th ACM international conference on Multimodal interaction}.\hskip 1em plus
  0.5em minus 0.4em\relax ACM, 2012, pp. 361--362.

\bibitem{Jin2015}
Q.~Jin, C.~Li, S.~Chen, and H.~Wu, ``Speech emotion recognition with acoustic
  and lexical features,'' in \emph{Acoustics, Speech and Signal Processing
  (ICASSP), 2015 IEEE International Conference on}.\hskip 1em plus 0.5em minus
  0.4em\relax IEEE, 2015, pp. 4749--4753.

\bibitem{Sarma2018}
M.~Sarma, P.~Ghahremani, D.~Povey, N.~K. Goel, K.~K. Sarma, and N.~Dehak,
  ``Emotion identification from raw speech signals using dnns,'' \emph{Proc.
  Interspeech 2018}, pp. 3097--3101, 2018.

\bibitem{Majumder2018}
N.~Majumder, D.~Hazarika, A.~Gelbukh, E.~Cambria, and S.~Poria, ``Multimodal
  sentiment analysis using hierarchical fusion with context modeling,''
  \emph{Knowledge-Based Systems}, 2018.

\bibitem{Hazarika2018}
D.~Hazarika, S.~Poria, A.~Zadeh, E.~Cambria, L.-P. Morency, and R.~Zimmermann,
  ``Conversational memory network for emotion recognition in dyadic dialogue
  videos,'' in \emph{Proceedings of the 2018 Conference of the North American
  Chapter of the Association for Computational Linguistics: Human Language
  Technologies, Volume 1 (Long Papers)}, vol.~1, 2018, pp. 2122--2132.

\bibitem{Boucenna2014}
S.~Boucenna, S.~Anzalone, E.~Tilmont, D.~Cohen, and M.~Chetouani, ``Learning of
  social signatures through imitation game between a robot and a human
  partner,'' \emph{IEEE Transactions on Autonomous Mental Development}, vol.~6,
  no.~3, pp. 213--225, 2014.

\bibitem{Ranieri2016}
C.~M. Ranieri, R.~A.~F. Romero, H.~Ferasoli~Filho \emph{et~al.}, ``A mobile
  virtual character with emotion-aware strategies for human-robot
  interaction,'' in \emph{International Conference on Advanced Cognitive
  Technologies and Applications, 8th}.\hskip 1em plus 0.5em minus 0.4em\relax
  International Academy, Research and Industry Association--IARIA, 2016.

\bibitem{Barros2017}
P.~Barros and S.~Wermter, ``A self-organizing model for affective memory,'' in
  \emph{Neural Networks (IJCNN), 2017 International Joint Conference on}.\hskip
  1em plus 0.5em minus 0.4em\relax IEEE, 2017, pp. 31--38.

\bibitem{Lawrence1989}
L.~I. Lin, ``{A} concordance correlation coefficient to evaluate
  reproducibility,'' \emph{Biometrics}, vol.~45, no.~1, pp. 255--268, Mar 1989.

\bibitem{Zheng2018}
Z.~Zheng, C.~Cao, X.~Chen, and G.~Xu, ``Multimodal emotion recognition for
  one-minute-gradual emotion challenge,'' \emph{arXiv preprint
  arXiv:1805.01060}, 2018.

\bibitem{Simonyan2014}
K.~Simonyan and A.~Zisserman, ``Very deep convolutional networks for
  large-scale image recognition,'' \emph{arXiv preprint arXiv:1409.1556}, 2014.

\bibitem{King2009}
D.~E. King, ``Dlib-ml: A machine learning toolkit,'' \emph{Journal of Machine
  Learning Research}, vol.~10, no. Jul, pp. 1755--1758, 2009.

\bibitem{Soundnet2016}
Y.~Aytar, C.~Vondrick, and A.~Torralba, ``Soundnet: Learning sound
  representations from unlabeled video,'' in \emph{Advances in Neural
  Information Processing Systems}, 2016, pp. 892--900.

\bibitem{lim2015recipe}
A.~Lim and H.~G. Okuno, ``A recipe for empathy,'' \emph{International Journal
  of Social Robotics}, vol.~7, no.~1, pp. 35--49, 2015.

\end{thebibliography}

\end{document}